# Social Network Community Detection Based on Textual Content Similarity and Sentimental Tendency

Jie Gao[1], Junping Du[1*], Yingxia Shao[1], Ang Li[1] and Zeli Guan[1]

[1] Beijing Key Laboratory of Intelligent Telecommunication Software and Multimedia,
School of Computer Science (National Pilot School of Software Engineering),
Beijing University of Posts and Telecommunications,
Beijing 100876, China

**Abstract.** Shared travel has gradually become one of the hot topics discussed on social networking platforms such as Micro Blog. In a timely manner, deeper network community detection on the evaluation content of shared travel in social networks can effectively conduct research and analysis on the public opinion orientation related to shared travel, which has great application prospects. The existing community detection algorithms generally measure the similarity of nodes in the network from the perspective of spatial distance. This paper proposes a Community detection algorithm based on Textual content Similarity and sentimental Tendency (CTST), considering the network structure and node attributes at the same time. The content similarity and sentimental tendency of network community users are taken as node attributes, and on this basis, an undirected weighted network is constructed for community detection. This paper conducts experiments with actual data and analyzes the experimental results. It is found that the modularity of the community detection results is high and the effect is good.

## 1 Introduction

With the development of the Internet and network communication, as a form of Internet application emerging in the Web2.0 era, electronic social networks such as Facebook, Twitter, WeChat, blogs, etc. have gradually become an indispensable social channel and way in the daily life of the public [1][2][3][4]. Expressing personal opinions and thoughts on the Internet and sharing personal daily life are the main forms of people's use of such electronic social networks. In this type of cyberspace, a large number of Internet users will continuously publish massive amounts of information on various topics, such as posts, pictures, videos, comments and likes. A large number of network users will have different information preferences due to their different knowledge systems, hobbies, and discussion topics. Users with similar preferences often form a community, and users in the same community act as network nodes to express similar opinions in the network space, so the connections within the community will be closer than the connections between different communities [5][6][7], which constitutes the so-called social network.

Micro Blog is one of the most commonly used social networking platforms for internet users in our country [8][9]. The research on the community structure of Micro Blog network is a hot issue in the research field of community detection. At the same time, the research on it helps to find the behavior rules of user groups in social networks [5], and can also realize the reasonable grouping of Micro Blog network users, and quickly lock in the target group according to different actual needs, which has important theoretical and practical significance [1][10]. The user posting and commenting mechanism of Micro Blog makes the user relationship abstract into an undirected network, in which the nodes of the network represent each user, and the relationship between users can be represented as an edge between the commenter and the poster of a blog post.

The existing traditional community detection algorithms generally interpret the results of community detection from the perspective of "distance", which only considers the topology of the network, that is, the topological relationship between nodes, while ignoring the attributes of nodes, making the results of community detection lack semantics and inconvenient to understand and explain [7][11][12][13]. In this paper, we propose a Community detection algorithm based on Textual content Similarity and sentimental Tendency (CTST), aiming at the lack of semantics in the results of some traditional community detection algorithms, and considering the network topology and node attributes at the same time [14].

Our contributions are summarized as follows:
- This paper proposes a social network Community detection algorithm based on Textual content Similarity and sentimental Tendency (CTST). We take the text content as a measure of similarity of node attributes, combine the network structure with node attributes, and abstract the network into an undirected weighted graph.
- In addition, we put forward the concept of "sentiment bias value" for the sentimental tendency of text content. We construct the sentiment vector based on the sentiment analysis of the text content, and combine the sentiment vectors of two different users to obtain the final composite sentiment vector. The composite sentiment vector is further transformed and calculated to obtain the sentiment bias value between two nodes in the network.
- We evaluate our algorithm on two real-world datasets, and experiments verify that our algorithm performs better on undirected weighted networks, with higher modularity and improved the effect of community detecting.

The rest of the paper is organized as follows. In section 2 we give an overview of the related work. In section 3 we present and analyze the algorithm. In section 4, we show experimental results on real-world datasets. Finally, we give conclusion in section 5.

## 2  Related Work

The concept of social network analysis was first proposed by the famous British anthropologist Alfred Radcliffe-Brown. In the research process of social network analysis, the clustering analysis of node clusters in the network is also called "graph clustering" [15][16], which is what we call "community detection". Sociological theory holds that members of an online community often have similar views and cognition on a certain topic or certain things. Then, when comments on a specific topic are generated, due to the commonalities among members of the network community, the entire community will show a certain degree of consistency for these comments [17][18][19]. Based on this, the consensus reached by community members on these topics has certain rules to follow [5][6][20], which helps us to study and discover the behavior rules of user groups in social networks.

 The research on the community structure of social networks has achieved important research progress in many application scenarios and has obtained many practical applications. With the continuous advancement of research on community structure detection algorithms in complex networks [21][22][23][24], various community detection algorithms have been continuously proposed and improved. Classic community detection algorithms mainly include graph segmentation-based methods, hierarchical clustering methods, modularity-based methods [25][26][27], and spectral clustering-based methods. The traditional method based on graph segmentation is to complete community detection by continuously deleting edges connecting different communities until there is no cutting edge in the network. The main idea of the hierarchical clustering method is to identify the vertex sets with high similarity, which can be divided into two types: aggregation algorithm and splitting algorithm according to the different identification process. The idea of the aggregation algorithm is to fuse nodes with high similarity through continuous iteration, the splitting algorithm is to find and delete the edges connecting different communities in the process of community detection until there is no connection between all different community structures. The most famous splitting algorithm is the GN algorithm proposed by Newman and Girvan in 2002 [6][28], which iteratively divides the structure of the community by continuously deleting the edge which has the largest Edge Betweenness Centrality (EBC) in the network. Based on the idea of GN algorithm, Newman proposed an improved GN algorithm, also known as Fast Newman algorithm [29], aiming at the shortcomings of GN algorithm in 2004. It is also in this algorithm that the concept of "Modularity" was first proposed. The basic idea of the method based on spectral clustering is [30][31]: to analyze the eigenvectors of the Laplacian operator of the social network. Since the eigenvectors corresponding to non-zero eigenvalues in the Laplacian matrix are approximately equal, it is considered that the corresponding nodes should be in the same community. The Fast Unfolding algorithm is another Modularity-based community detection algorithm proposed by Blondel and other scholars in 2008 based on the idea of the Fast Newman algorithm [27]. The algorithm focuses on the concept of "community folding", and iteratively analyzes the entire community that has been discovered as a new node, thereby forming a community structure with a hierarchical structure. Many domestic scholars have also made a lot of contributions to the research field of community detection. Z XIE et al. optimized the Fast Unfolding algorithm and proposed a community detection algorithm with lower time complexity [32]. YW Jiang, CY Jia and other scholars used the Jaccard distance formula to measure the similarity between nodes in a social network [33], and used K-means, hierarchical clustering and other clustering methods to conduct community detection in complex networks, with good results.

Most of the traditional community detection algorithms measure the similarity of nodes from the perspective of spatial distance, and only consider the structure of the network, but ignore the attributes between nodes, which makes the results lack of semantics and inconvenient to interpret. In this paper, based on the simultaneous consideration of the two aspects, we propose the CTST model, which uses text similarity and sentiment tendency as a measure of the similarity of node attributes, which improves the problem of sparse attribute data to a certain extent and divides the community. The results endow certain semantics and interpretability, and improve the effect of community detection.

## 3   Methodology

In this section, we first describe the overall situation of the CTST model, and give the mathematical calculation of the content similarity and propose the concept of the sentiment bias value, and then show the details about our algorithm.

### 3.1   Overview

Fig. 1 shows the overall structure of our CTST model. It can be seen that the CTST model is generally divided into two routes to process the user's comment text. On the one hand, the upper part of Fig. 1 shows the content similarity calculation route. First, the CTST model performs word segmentation on textual content, then calculate the $tf - idf$ value of the word entry after word segmentation, and extract the text information eigenvector based on this. Then use the formulas we proposed to calculate the cosine similarity of the two eigenvectors and use it as a measure of the similarity between the two user nodes. On the other hand, in the sentiment bias value calculation part shown in the lower part of Fig. 1, the model first constructs the basic sentiment vectors of the two users, and then combines them. After a series of transformations, the sentiment bias value is calculated. Finally, the model will synthesize the results of two parts and construct the undirected weighted network, and on this basis, community detection is carried out on the network [34][35].

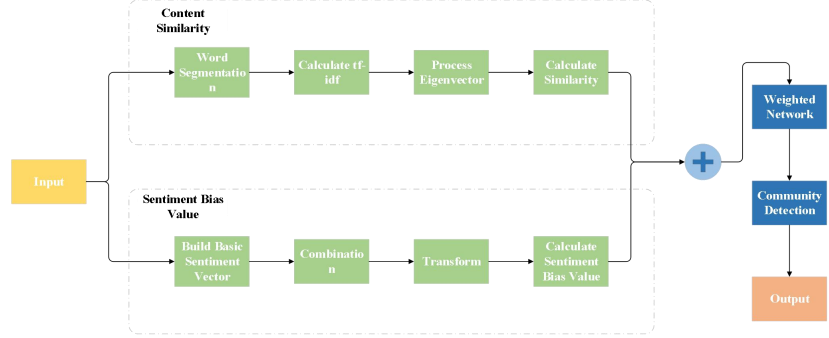

**Fig. 1.** The overall logical architecture of our algorithm. It first consists of two separate parts. On the left, it calculates the *Content Similarity*, and it calculates the *Sentiment Bias Value* on the right. These two values are then processed into weights, and *Community Detection* is performed on the *Weighted Network*.

### 3.2  Content Similarity Calculation

First, we perform word segmentation on the text content of the comment information of the user nodes, and then calculate the $tf-idf$ value of the entry. The $tf$ value represents the frequency of a certain word in the text. Its calculation formula is shown in Formula (1), where the numerator $n_{i,j}$ represents the frequency of a word in the text $d_j$, and the denominator represents the total number of words in the entire text, and the $tf$ value is used to represent the importance of a word in the entire text content.

$$tf_{i,j} = \frac{n_{i,j}}{\sum_k n_{k,j}} \quad . \tag{1}$$

$idf$ represents the reverse document frequency, and its calculation formula is shown in Formula (2). The numerator $|D|$ represents the total number of documents in the corpus, and the denominator $\{j : t_i \in d_j\}$ represents the number of documents containing the word $t_i$, and then take logarithm of the quotient to get $idf$ value.

$$idf_{i,j} = \log\frac{|D|}{\{j : t_i \in d_j\}} \quad . \tag{2}$$

Multiply the obtained $tf$ value by the $idf$ value, as shown in Formula (3), to get the $tf-idf$ value of the term:

$$tf - idf_{i,j} = tf_{i,j} \times idf_{i,j} \quad . \tag{3}$$

After the $tf - idf$ value of the entry is obtained, the user text information is processed to obtain the information text eigenvector between any two network user nodes $V1$ and $V2$, and the cosine similarity between the two eigenvectors is calculated as the similarity measure $s$ between the two user nodes. The specific calculations are shown in Formula (4) and Formula (5). Among them, $tf - idf_{V1}$ represents the text eigenvectors of user node $V1$, and $tf - idf_{V2}$ represents the text eigenvectors of user node $V2$.

$$s(V1, V2) = \cos(tf - idf_{V1}, tf - idf_{V2}) \quad . \tag{4}$$

$$\cos \theta = \frac{\sum_{i=1}^{n}(V1_i \times V2_i)}{\sqrt{\sum_{i=1}^{n} V1_i^2} \times \sqrt{\sum_{i=1}^{n} V2_i^2}} \quad . \tag{5}$$

### 3.3  Sentiment Bias Value

Each user node in the network has corresponding text information content. Correspondingly, the text information contains the user's opinions, emotions and attitudes about the topic. There are two kinds of traditional sentiment analysis methods, one is based on sentiment dictionary and the other is based on machine learning [36][37].

Micro Blog has a mechanism for users to post comments. From the perspective of emotion, the poster expresses his emotions in the post, and this emotion will affect the emotional expression of the comment users who reply to the blog post to some extent. From the perspective of user relationship, these two users have emotional connection in addition to the connection caused by mutual comments, which we call "sentiment bias". Based on this, we first construct a polar coordinate system. In the polar coordinate system, the emotional state of each user is represented by an sentiment vector $e_i$, which is represented by Formula (6):

$$e_i = (\rho_i, \omega_i) \quad . \tag{6}$$

In the above formula, $\rho_i$ and $\omega_i$ represent the emotional intensity and the corresponding weight assigned to it, respectively. Each piece of text information has different degrees of emotional tendency, so the polar diameter $\rho_i \in [0,1]$ is introduced to describe the intensity of emotion, that is, the emotional score. The two sentiment vectors can be transformed and combined to produce a composite sentiment vector.

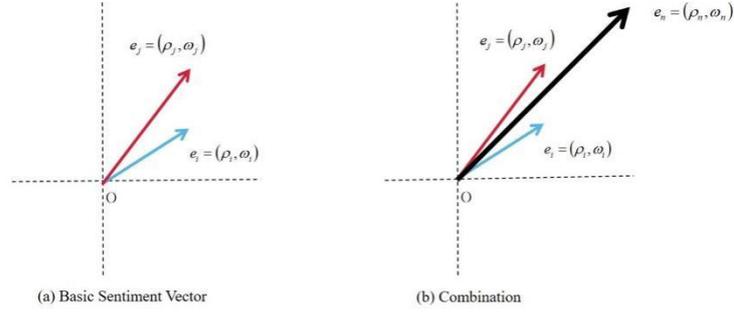

**Fig. 2.** The basic sentiment vectors of the two users $e_i$ and $e_j$ are combined to obtain a composite sentiment vector $e_n$.

The composite sentiment vector can be further transformed and calculated to obtain the final sentiment bias value. The calculation of the sentiment bias value is shown in Formula (7).

$$sv = \rho_n \times \omega_n \ . \tag{7}$$

### 3.4 Weighted Network

We map network users to the graph. Suppose that $V = \{V_1, V_2, \ldots, V_i, \ldots, V_n\}$ is the set of network user nodes, where $(V_i, V_j)$ represents an edge between two nodes. $W_{ij}$ is the weight of the edge, which consists of the text content similarity value $s$ and the sentiment bias value between user $V_i$ and user $V_j$, as shown in Formula (8). Then $WG(V, E, W)$ is an undirected weighted graph with $V$ as the user node set, $E \in \{(V_i, V_j) | V_i, V_j \in V\}$ as the edge set, and $W = \{W_{ij} : (V_i, V_j) \in E\}$ as the weight set. On this basis, the algorithm is used to divide the graph into communities, and the community detection result of the social network user communities is obtained.

$$W_{ij} = s \times 0.5 + sv \times 0.5 \ . \tag{8}$$

## 4  Experiments

### 4.1  Datasets

In recent years, the development of shared travel relying on the mobile Internet has become one of the hot topics discussed on social networking platforms such as Micro Blog. At the same time, the evaluation content of the shared travel mode by the majority of social network users is also likely to form a public opinion orientation that affects the supply and demand relationship of shared travel [38]. Therefore, a deeper social network community detection based on the Micro Blog social network and the comment content of shared travel can help us more effectively study the characteristics and laws of user groups with different preferences, so as to analyze the public opinion of the comment content on shared travel. It has good research value and practical significance [39].

Since there is no public available Chinese sentiment analysis dataset related to the comment content of "shared travel", this paper uses the Micro Blog platform as the data collection platform, crawls the links of some discussion blog posts under the relevant entries of "shared travel" on Micro Blog, and processes the data accordingly. In the actual experiment, 1693 pieces of blog data were finally obtained, and an undirected weighted network was established according to the method described above. In addition, in order to demonstrate the advantages of the algorithm proposed in this paper on undirected weighted networks, we also conducted comparative experiments on the classic public dataset named Zachary's Karate Club, which is an undirected and unweighted network with 34 nodes and 156 edges.

### 4.2  Experimental Procedure

We first processed the data and obtained 85 actual users from the 1693 blog post data. According to the method described above, an undirected graph with 85 nodes and 396 edges is established. The user's text information is calculated according to the method described in Section 3.2, and the similarity matrix is obtained as shown in Table 1.

Table 1.  Textual content similarity matrix.

| Nodes | 1 | 2 | 3 | ... | 85 |
|---|---|---|---|---|---|
| 1 | 0 | 0.24 | 0.53 | ... | 0.05 |
| 2 | 0.24 | 0 | 0.62 | ... | 0.12 |
| 3 | 0.53 | 0.62 | 0 | ... | 0.33 |
| ... | ... | ... | ... | ... | ... |
| 85 | 0.05 | 0.12 | 0.33 | ... | 0 |

Then, according to the method described in Section 3.3, the sentiment bias value matrix is calculated and shown in Table 2.

**Table 2.** Sentiment bias value matrix.

| Nodes | 1 | 2 | 3 | ... | 85 |
|---|---|---|---|---|---|
| 1 | 0 | 0.37 | 0.6 | ... | 0.22 |
| 2 | 0.37 | 0 | 0.81 | ... | 0.25 |
| 3 | 0.6 | 0.81 | 0 | ... | 0.59 |
| ... | ... | ... | ... | ... | ... |
| 85 | 0.22 | 0.25 | 0.59 | ... | 0 |

Finally, a weighted network is established according to the contents described in Section 3.4. And based on the above results, we add the weighted edges to the undirected network and perform further community detection on the basis of the network. We divide the process of community detection into two steps. The first step is to select appropriate central nodes as the center of the clusters, and the second step is to determine whether to add a node to the cluster by checking the correlation of adjacent nodes. During the experiment, we set the number of initial central nodes to 2, 3, and 4 respectively to divide the undirected weighted network into communities.

### 4.3 Experimental Results

We use the modularity to judge the result of community detection, which is a value between [0, 1]. The larger the modularity, the tighter the internal connection of the community in the result of community detection, and the better the connectivity. Its specific calculation is shown in Formula (9). In the following formula, $L_n$ represents the edge within the community, $L$ represents the number of all edges in the network, and $D_n$ represents the sum of the degrees of all nodes in the community.

$$Q = \sum_{n=1}^{m} [\frac{L_n}{L} - \left(\frac{D_n}{2L}\right)^2 ] \ . \quad (9)$$

For the three experiments on our dataset, we calculated their modularity separately, and the results are shown in Table 3.

**Table 3.** Modularity comparison.

| Number of initial central nodes | modularity |
|---|---|
| 2 | 0.298 |
| 3 | 0.329 |
| 4 | 0.443 |

It can be seen from Table 3 that the modularity increases with the increase of the number of initial central nodes, and the effect of community detection is best when the number of initial central nodes is 4. In addition, we also compared the algorithm on the Zachary's Karate Club dataset, and found that the modularity of the CTST model on the undirected weighted network is higher than that on the undirected unweighted network, that is, the community detection effect on the undirected weighted graph is better.

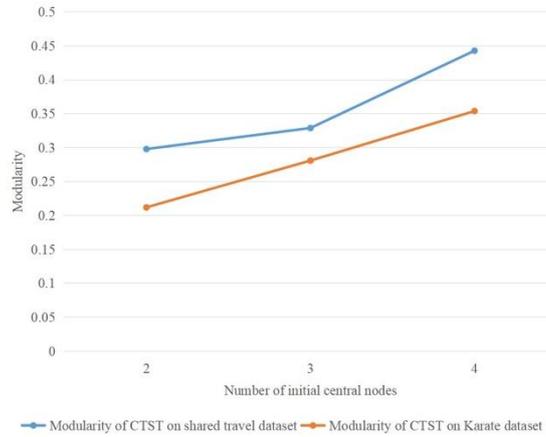

**Fig. 3.** The comparison results of our CTST model on *shared travel dataset* and *Karate dataset*.

## 5  Conclusion

In this paper, we propose a social network Community detection algorithm based on Textual content Similarity and sentimental Tendency (CTST). We measure the similarity of user nodes by using the similarity and sentimental tendency of the relevant comment texts of users on the topic of "shared travel" in the social network as node attributes. It is worth mentioning that we propose the concept of sentiment bias value as a numerical expression of sentiment bias in the emotional orientation part of the text content. Experiments have shown that our CTST model, which takes advantage of the rich information contained in the text content and considers the network structure and node attributes, can not only add semantics to the results of community detection, but also describe similar users in the same community. It is easier to understand and explain, and to a certain extent solves the problem of sparse attribute data that may be caused by only considering a single situation. It makes the user nodes within the community more similar and more stable, and improves the quality of community detection. In future work, we will focus on researching the results of community detection based on shared travel-related comment content, and analyze its impact on public opinion orientation, and based on this, we will realize online public opinion monitoring and real-time early warning.